# Data Combination for Problem-solving: A Case of an Open Data Exchange Platform

Teruaki Hayashi[1], Hiroki Sakaji[1], Hiroyasu Matsushima[2],
Yoshiaki Fukami[3], Takumi Shimizu[4], Yukio Ohsawa[1]

[1]Department of Systems Innovation, School of Engineering, The University of Tokyo, Japan
[2]Center for Data Science Education and Research, Shiga University, Japan
[3]Graduate school of Media and Governance, Keio University/ Department of Management, Faculty of Economics, Gakushuin University
[4]Graduate School of Business and Finance, Waseda University, Japan

**Abstract**

In recent years, rather than enclosing data within a single organization, exchanging and combining data from different domains has become an emerging practice. Many studies have discussed the economic and utility value of data and data exchange, but the characteristics of data that contribute to problem solving through data combination have not been fully understood. In big data and interdisciplinary data combinations, large-scale data with many variables are expected to be used, and value is expected to be created by combining data as much as possible. In this study, we conduct three experiments to investigate the characteristics of data, focusing on the relationships between data combinations and variables in each dataset, using empirical data shared by the local government. The results indicate that even datasets that have a few variables are frequently used to propose solutions for problem solving. Moreover, we found that even if the datasets in the solution do not have common variables, there are some well-established solutions to the problems. The findings of this study shed light on mechanisms behind data combination for problem-solving involving multiple datasets and variables.

## 1 Introduction

In recent years, the practice of creating new businesses and adding value to existing services by exchanging and combining data from different domains has been emerging. Rather than using data encompassed within a single organization, other companies have increasingly been publishing data for use, and third-party data are increasingly being combined [1, 2]. Platform businesses that develop a marketplace for exchanging different types of data, such as open data and sensitive data owned by individuals and companies, have been launched, thereby forming a business ecosystem [3–5]. In addition to the expectations for data exchange, in the global trend of big data, open data, and data combination among different domains, large-scale data are expected to be used [6, 7].

Under such circumstances, it is believed that combining large amounts of multivariable data can create valuable solutions and services. In addition, such big data may be expected to be "universal data" that can contribute to solving all the problems. However, Bollier pointed out that although new values are expected to be derived from data combinations, heterogeneous data combinations make objective interpretation difficult [8]. Boyd and Crawford also stated that the volume of data is meaningless when the meaning of data is not considered, and it is important to understand the value of small amounts of data stored in various domains [9]. Although many studies have recognized the advantages and value of big data, they have also identified the limitations and issues of the secondary use and aggregation of big data [10, 11]. In addition, the characteristics of data that contribute to problem solving through data combination–the combinability or co-occurrence patterns of data–have not been fully understood.

To address this important issue, this study attempts to understand the relationships between the expectation of usage and the combinability of data. We analyze the characteristics of the data that constitute a solution in open data utilization considering the number of combined data with variables. At present, the evaluation criteria for open data have not been established. The characteristics of data that are combinable and lead to useful solutions have not been adequately addressed. The chief contribution of this study is the analysis of the combinability of relevant data to the solutions using empirical data provided by a local government.

The remainder of this paper is organized as follows. In Section 2, we discuss the issues addressed in our study and present some related works. In Section 3, we present the experimental details of the datasets and the analysis method. In Section 4, we discuss the results and limitations of the current approach and areas for further study. Finally, we conclude the paper in Section 5.

## 2 Research Questions and Related Works

Presently, the environment and infrastructure for data exchange and utilization are being rapidly developed, and a type of ecosystem related to data is being formed. Boisot and Canals argued that data, information, and knowledge are distinct types of economic goods, each of which has a specific type of utility [12]. Mergers and acquisitions have been actively conducted with certain expectations in terms of the value of the data assets [13], and some firms have opened their application programming interfaces (APIs) to sell their data resources. Amazon, for example, has opened the API of product databases based on their marketing strategy [14]. Such activities being performed by many companies to create new business models through API disclosure constitute the API economy [15]. The financial sector is also aggressive in creating business models by exposing APIs [16]. Data exchange and combination through APIs now considerably affects the economy.

Open data–machine-readable information, particularly government data–are another component of the data exchange ecosystem [7, 17, 18]. Although the term "open data" often refers to public sector information, data providers are not limited to governments. Some open data are provided by private organizations to revitalize the economy and create new businesses [18]. For example, financial authorities in many countries require companies to disclose their financial status using XBRL (eXtensible Business Reporting Language), which is a markup language used for corporate electronic accounting reporting [19]. Business models that use open data from aggregators, brokers, and service providers have been reported [20, 21], and Zimmermann and Pucihar highlighted the value of open data sources [22].

The economic and utility values of data have been discussed in many studies, but the debate over what types of data are valuable remains ongoing. Without exception, the method for data valuation has not been established in open data. Moreover, the characteristics of combinable data have not been fully clarified. Some research has argued the importance of a collaborative environment to support businesses based on open data [20, 23], but there has been no discussion about the relationships among diversified open data resources.

To tackle this challenging issue, we focus on variables as the characteristics of data combination in this study. A variable is a logical set of data attributes. Data attributes are important features that can be used to understand the structure and granularity of data. For example, streetlight data might contain variables such as "latitude," "longitude," "lump type," "luminous flux," and so on, and "population," "ward name," "age," and "sex" are likely variables included in demographic data. Variables are important for discussing characteristics such as connectivity with other data [24, 25]. In interdisciplinary data combination, there is an expectation that highly used data will include large-scale data with many variables. Given that expectation, are the data with fewer variables less likely to be used? We analyzed the relationship between the number of variables in the data and the frequency of data usage as research question #1.

Variables are not the only important aspect in data utilization. In data combination, the creation of value by combining data from different domains is highly expected. The question here is whether combining a large number of data will increase the value. To discuss this, we analyze the number and distribution of combined data for problem solving as research question #2.

The third research question is concerned with the context of data combination, where context is a solution to a problem. Even with the same map data, for example, the context is different between the solution "creating a hazard map of the area where you live in combination with disaster information" and "understanding city congestion by overlaying a map with people flow data." The combination of data and the number of combinations required to achieve the solution may vary depending on the data usage context. Based on this assumption, we analyze the combination types of data with the usage context of the solution as research question #3.

## 3 Experimental Details

In this study, we aimed to investigate the characteristics of the data and solutions that contribute to problem solving, while focusing on the relationships between data combinations and variables in open data exchange. However, although open data are a publicly available information source on the Web, knowledge of how to use them for a certain purpose is not common. Therefore, in this study, we used a database in which the information on datasets and how to use them are stored as data jackets (DJs) that contain structured knowledge regarding data utilization. DJ is a metadata format used to describe the summary information of datasets. Even if the datasets themselves cannot be widely published owing to sensitivity of the data, by sharing the summary information, it is possible to read and understand the characteristics and their structure [26]. Table 1 presents an example of a DJ on "event information" that was stored in the knowledge base we used in the experiment.

The two primary advantages of using the knowledge base with DJs are the descriptions of variables and linkages with knowledge elements of problem solving. In the DJs, information on variables is stored as variable labels, written in natural language. The number of variable labels in each dataset varies, which is useful for verifying the research questions of this study. The other advantage is that the knowledge base stores not only information on datasets but also the dataset usage contexts as solutions and requirements. The solution summarizes dataset utilization with combined data, and the requirements are the needs written in natural language. The knowledge base is created by combining the following two equations with binary predicate logic [27], where Eq. (1) formulates the relationship such that a certain solution satisfies a requirement, and Eq. (2) formulates that a combination of the DJs generates a solution.

$$\textbf{satisfy}(solution, requirement) \quad (1)$$

$$\textbf{combine}(solution, DJ) \quad (2)$$

Table 1: Example of a DJ

| Data name | Event information |
|---|---|
| ID | 3502 |
| Data outline | This is a dataset obtained using a search tool to identify events. The tool provides information on when, where, and what kinds of events will be held. |
| Data type | Text, image, numerical value |
| Variable labels | Event name, date, event type, target, venue, participation fee, capacity, organizer, contact information |

To investigate the research questions, we used datasets available from a platform provided by the Institute of Administrative Information Systems (IAIS)[1], which comprised 623 DJs, 158 solutions, and 273 requirements. The DJs on the IAIS data platform include all available open data for Yokohama City[2] and part of the open data for Kawasaki City[3], both of which are cities in Kanagawa Prefecture, Japan. It should be noted that the original datasets for Yokohama City have been completely moved to the new data catalog site in 2020, and some datasets are difficult to identify with the DJs we used in the experiment. Although there were 676 DJs in total on the IAIS platform, we integrated the DJs with the same dataset names and variables and used them without duplication. Moreover, even if the dataset names were the same, those with different variables were treated as different DJs. In addition, we manually corrected the mistakes of the variable delimiters and typographical errors.

The solutions and requirements in the data were created using DJs at workshops of Innovators Marketplace on DJs (IMDJ [28]) held in Yokohama and Kawasaki. A hundred people in Yokohama City and 29 in Kawasaki City–comprising citizens, city office workers, and data utilization professionals–participated in the workshops. At the workshop, participants presented their problems and social issues as requirements for the first 15 min. Then, the participants proposed problem solving methods as solutions for satisfying the requirements by the combination of data written in DJs. In addition, the participants could supplement additional datasets for creating solutions during the discussion and evaluate the solutions that meet their requirements or merit implementation using an imaginary purchasing budget provided to them. However, it should be noted that information on additional datasets and solutions evaluated based on the participants' purchasing budget were not stored on the IAIS platform, and we did not use them in our analysis. The workshop proceeded for approximately 90 min. For more details regarding the rules followed in the workshop, see references [28, 29].

The data (DJs, requirements, solutions, and their relationships) were created in workshops conducted under the theme of "creating Yokohama, a city that can be enjoyed with children" in Yokohama City and "community formation for intergenerational exchange" in Kawasaki City. Both themes utilize open data from local governments to express the opinions of citizens and propose solutions to their problems. From a global perspective, the themes concern public participation and welfare, and we treated the quality of data combinations in the solutions created in the two different workshops as equivalent in this study. As all DJs, requirements, and solutions were written in Japanese, the analysis was conducted in Japanese and translated into English when the present paper was written.

---

[1] https://djp.iais.or.jp/s/djplatform
[2] https://data.city.yokohama.lg.jp/
[3] https://www.city.kawasaki.jp/shisei/category/51-7-0-0-0-0-0-0-0.html

## 4 Results and Discussion

### 4.1 Variables and the frequency of dataset use

Figure 1 presents the frequencies of use of the 43 datasets to create the solution and the number of variables that each dataset comprise. The size of each dot indicates the number of occurrences of datasets with the same number variables against the frequency of use (maximum: 3 times, minimum: 1 time). For example, three datasets "list of local Terakoya[4] projects," "the location of the stations which installed the elevators," and "list of facilities where we can breastfeed and change diapers for babies" have three variables each and were used three times. Therefore, the number of occurrences (the size of the dot) was three. As can be observed in Fig. 1, we could not find a correlation between the number of variables in each dataset and the frequency of dataset usage ($r = -0.0291$). Thus, it can be said that datasets with large numbers of variables are not always used to create solutions.

By contrast, datasets with a small number of variables appeared to be used more frequently. For example, the most frequently used dataset was "event information," which was used eight times to create solutions and contained only nine variables (Fig. 1). In addition, of the 131 datasets used, including duplication, the ratio of datasets with 1 to 10 variables was high at 74%. Figure 2 shows the top 15 datasets used to create solutions. The numbers in parentheses indicate the number of variables included. Although the number of variables ranges from 4 to 21, the datasets that contribute to creating solutions contain fewer variables.

### 4.2 Number of data to create solutions

Next, we analyzed and compared the number of data combined to create a solution, as shown in Fig. 3. Ninety-nine solutions do not use data or have no links to data in any dataset; therefore, we targeted the remaining 59 solutions that use data. The solution that combined the most datasets was "providing the happiness ranking of the children by facilities which can take care of children;" this solution used seven datasets: "unlicensed childcare facility," "list of children- and baby-friendly restaurants," "list of infant care facilities," "playground for children," "data of local child-rearing support centers," "event data for infant and children," and "list of hospitals with daycare centers or children's play area." This solution used a high number of administrative datasets on children and babies. However, most solutions consisted of a small number of variables (average: 2.22). It is worth noting that 17 solutions that use only one dataset have been proposed to satisfy the requirements. This result suggests that the solutions are not necessarily composed of a large number of datasets, but a few dataset combinations are enough to establish the solutions.

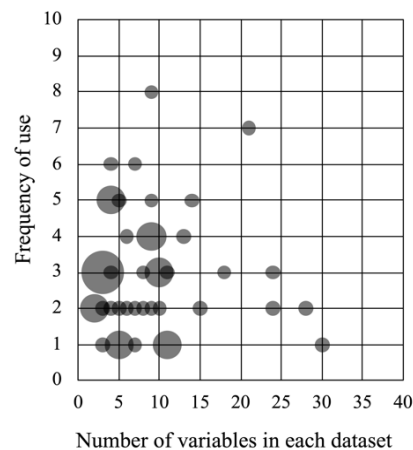

Figure 1: Number of variables in each dataset and the frequency of use.

---

[4] Terakoya is the local private elementary school in Japan that originates from the temple schools of the Edo era.

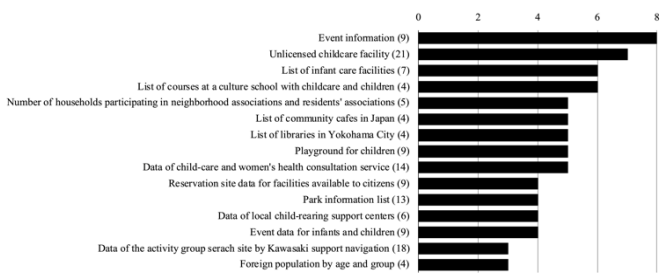

Figure 2: Top 15 frequently used data.

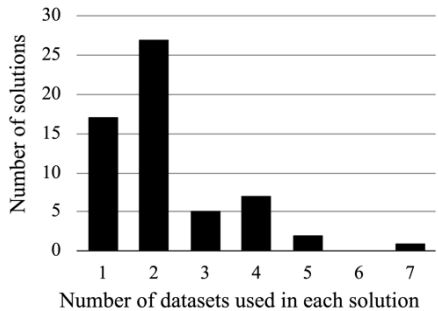

Figure 3: Number of data to create solutions.

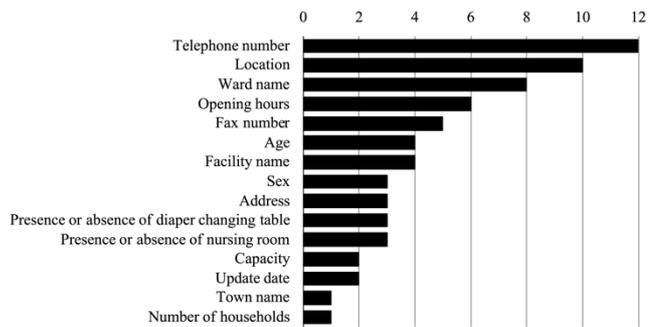

Figure 4: Top 15 variables in the datasets used for solutions.

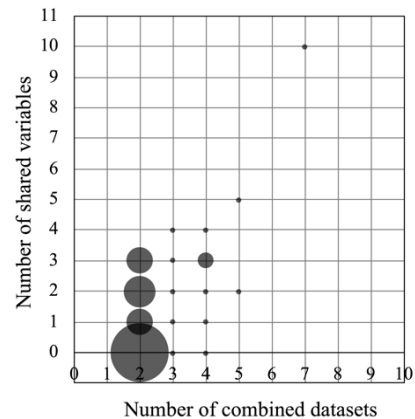

Figure 5: Numbers of combined datasets and shared variables.

### 4.3 Shared variables and the combinability of data

Forty-two of the 59 solutions were created using two or more datasets. Of these solutions, we found that 13 were composed of datasets that did not have shared variables, and 29 consisted of datasets that had one or more common variables. This accounts for approximately 70% of the total, and it can be said that many solutions are created by combining datasets that contain common variables. Figure 4 presents the top 15 frequently appearing variables required to create the solutions. Many solutions share the following variables: "telephone number," "location," "ward name," and "opening hours." For the contexts of child rearing and local community building, for example, solutions such as "setting up a shared office for those raising children" and "establishing a reservation service for facilities where you can find friends to exercise with your children" have been proposed. For the development of services rooted in the community, "location," "telephone number," and "opening hours" of the facilities may be essential variables across datasets. By contrast, many variables such as "ward name," "age," and "sex" were shared in "conducting an international exchange conversation class" and "holding a grandma's wisdom cafe." In holding the events, it is necessary to share variables regarding the areas covered by each event, target age, and gender among the datasets.

Figure 5 presents a comparison of the number of combined datasets with the number of shared variables, where the target was the solutions that used two or more datasets. The size of the dots represents the frequency of the solutions (maximum: 11, minimum: 1). Most of the solutions have been created by combining 2–4 datasets, and the number of shared variables in the combination has varied from 0 to 4. In other words, it can be said that a solution with a small number of combined datasets does not always have a small number of shared variables, and similarly, a solution with a large number of combined datasets does not have a large number of shared variables. It is interesting to note that there are solutions with extremely large numbers of combined datasets and shared variables, which were created by combining seven datasets with 10 shared variables. The solution was "providing the happiness ranking of the children by facilities which can take care of children," and it used multiple facility datasets and event datasets for children. Therefore, it is necessary to share many variables such as "address," "facility name," "presence or absence of car parking," "presence or absence of nursing room," and their "fee."

Does a solution exist that consists of datasets with no common variables? The solution of "holding Terakoya for international exchange" was proposed to arrange a private elementary school for international exchange in a shopping district or at a cafe by gauging the necessity for international exchange from foreign residents. In this solution, "foreign population by age and group," "list of community cafes in Japan," "list of shopping arcades," and "list of local Terakoya businesses" were used, but they had no common variables. This solution did not solve the problem by combining datasets from common variables in parallel; instead, it processed the datasets serially according to realization steps. A realization step is a step-by-step task to achieve a solution. The datasets are processed in each step as required, and finally, the solution is achieved (Fig. 6(b)). Therefore, these solutions do not need to have common variables. In the solution "counseling to respond to the anxiety of advanced maternal age" the following three datasets were used: "average age at birth of mothers," "data of child-care and women's health consultation service," and "list of hospitals with daycare centers or children's play area." This solution also used the datasets based on realization steps and did not integrate the datasets. First, the solution determines the counseling demands using the average childbirth age in the area, and second, it decides the content of child-care/health counseling. Finally, the solution achieves the counseling at a hospital with a daycare center. By contrast, a dataset with shared variables is presented in Fig. 6(a). The integrated dataset is created by combining multiple datasets with common variables. Then, the dataset is analyzed, processed, or incorporated into the system to generate a solution. All solutions created using the datasets can be divided into either of these two combination types.

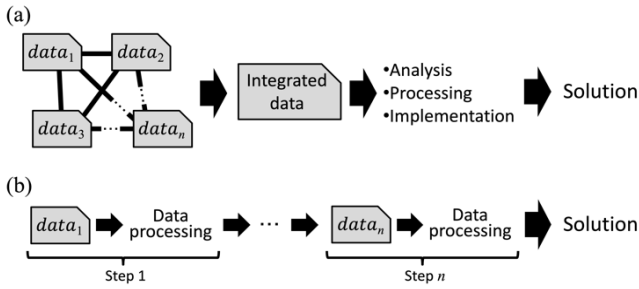

Figure 6: Data combination types.

Finally, we examine how the commonality of variables is related to dataset combinations and solution creation. Of the 43 datasets used in the solutions, 263 dataset pairs shared variables. The number of dataset pairs used to create useful solutions was 65, of which 38 dataset pairs shared variables. In other words, there were 225 dataset pairs that had never been combined, even though they shared many variables. Hence, it can be said that the commonality of variables is not a prerequisite for creating a solution; instead, the datasets to be combined are selected according to the context of the problem or the solution to be achieved.

From the analysis results, it was found that solutions may have a various number of combined datasets, and the datasets used to create the solutions do not necessarily share many variables. Furthermore, even if the datasets in the solution do not have common variables, there are some solutions that are adequate for the problems. The result suggests that variable sharing is not the only important factor; the context–how to combine the datasets to generate a solution to solve the problem–is essential for dataset combination.

### 4.4 Summary of discussion

We conducted three experiments to answer the research questions. The first result revealed that the datasets required for problem solving, that is, solution creation via dataset combination, do not necessarily have many variables (research question #1). In other words, even datasets with a few variables can be used effectively to solve problems. The results of the second experiment demonstrated that the solution was not necessarily created by combining a large number of datasets (research question #2). Some solutions were created by combining one or two datasets; hence, it is not necessary to combine a large number of datasets to create solutions. In addition, we found that the number of combined datasets varied, and some solutions were sufficiently established even if they did not have common variables (research question #3). We found that the combination of datasets can be divided into two types: integrated dataset creation using common variables and step-by-step usage of data.

### 4.5 Limitation and future work

In this study, we analyzed the data utilization knowledge base from the viewpoint of open data exchange. However, there are some limitations owing to the lack of data in the IAIS data platform. In future research, it will be necessary to focus on the following three points and clarify the mechanism of contexts of data use and data valuation.

The first limitation is the causal relationship between solution creation and the number of variables. Experiments have shown that datasets with a large number of variables are not necessarily used many times to create a solution. Then, are datasets with fewer variables used more often to create solutions? The answer is probably NO. It is known that the distribution of the number of variables in the datasets on the data exchange platform is the power distribution [25]. In other words, datasets with a small number of variables account for most of the data population, and datasets with many variables occur rarely. In fact, the datasets used in this study are also biased toward datasets with fewer variables, which makes it possible that the distribution of the datasets with a small number of variables may have affected the result. Therefore, it was not possible to compare and examine the datasets we used in the experiment. For further study, it will be effective to use the knowledge base of the DJ store [27] and Web IMDJ [29]. These knowledge bases store the results of multiple past workshops for data utilization. Future studies will benefit by using these datasets to deepen the discussion of the analysis.

The second is the evaluation criteria of the datasets and solutions. In this study, we evaluated and discussed the usage expectation based on the number of dataset usages and variables for creating solutions. However, solutions have to be evaluated based on feasibility or usefulness or even the data owners' price for the data, which may influence the solution value. To solve this problem, the knowledge bases of the DJ Store and Web IMDJ might be helpful. In the future, we will evaluate solutions using an imaginary purchasing budget in the workshop and payment information stored in the databases. As a result, it will be possible to discuss the value of data and the contexts of data usage.

The third limitation is the sharing condition of the data. The Yokohama and Kawasaki data that we used in this study were all open data. The data marketplace includes not only the shareable government data but also treats the sensitive data from companies and individuals with multiple stakeholders [30–32]. It is known that sensitive data and shareable data have unique characteristics in terms of variables and their connectivity [25]. The contexts of data utilization, how they are used, and how often they are used may depend on the sharing conditions of the data. Future studies are required to analyze the availability of heterogeneous data and their sharing conditions.

## 5 Conclusion

In this study, we analyzed how different types of data are used to build an open data solution from the viewpoint of the number of combined datasets with variables. The results of the experiments indicated that many solutions have been proposed, and it is not always necessary to combine numerous datasets. Furthermore, it was suggested that a solution can be created even with datasets that have a small number of variables. It was also found that the combination of datasets is not limited to a parallel combination because of the commonality of variables; there is a combination type in which datasets are combined in series based on the realization steps of the solution, which do not required shared variables. In big data and interdisciplinary data combination, it is expected that large-scale data with many variables will be used, and value will be created by combining data as much as possible. We believe that the findings of this study will balance expectations for solutions involving multiple dataset combinations and numerous variables.

However, the data exchange ecosystem still lacks observable events–the value of data, the transaction of data, communication logs among stakeholders, and so forth– which makes it difficult to obtain sufficient data to test hypotheses. In the future, as mentioned in Section 4.5, it will be necessary to apply our analysis to other data to clarify the data ecosystem where innovation occurs through heterogeneous data exchange.

## Acknowledgments

This study is supported by KAKENHI JP20H02384 and the Artificial Intelligence Research Promotion Foundation. We would like to thank the Institute of Administrative Information Systems for sharing their data.
## References

[1] Balazinska, M., Howe, B., Suciu, D. (2011). Data markets in the cloud: An opportunity for the database community. *Proceedings of the VLDB Endowment*, 4(12), 1482–1485.

[2] Stahl, F., Schomm, F., Vossen, G. (2014). Data marketplaces: An emerging species. *Frontiers in Artificial Intelligence and Applications*, 145–158.

[3] Liang, F., Yu, W., An, D., Yang, Q., Fu, X., Zhao, W. (2018). A survey on big data market: Pricing, trading and protection. *IEEE Access*, 6, 15132–15154.

[4] Hayashi, T., Ishimura, G., Ohsawa, Y. (2020). Description framework for stakeholder-centric value chain of data to understand data exchange ecosystem. Principle and practice of data and knowledge acquisition workshop (in press).

[5] Fernandez, R.C., Subramaniam, P., Franklin, M.J. (2020). Data market platforms: Trading data assets to solve data problems. *Proceedings of the VLDB Endowment*, 13(12), 1933–1947.

[6] Manyika, J., Chui, M., Brown, B., Bughin, J., Dobbs, R., Roxburgh, C., Byers, H.A. (2011). Big data: The next frontier for innovation, competition and productivity. McKinsey Global Institute.

[7] Manyika, J., Chui, M., Groves, P., Farrell, D., Kuiken, S.V., Doshi, E.A. (2013). Opendata: Unlocking innovation and performance with liquid information. McKinsey Global Institute.

[8] Bollier, D. (2010). The promise and peril of big data. Communications and Society Program. The Aspen Institute, Washington, DC.

[9] Boyd, D., Crawford, K. (2012). Critical questions for big data. *Information, Communication and Society*, 15(5), 662–679.

[10] Ellram, L.M., Tate, W.L. (2016). The use of secondary data in purchasing and supply management (P/SM) research. *Journal of Purchasing and Supply Management*, 22(4), 250–254.

[11] Gregory, A., Halff, G. (2020). The damage done by big data-driven public relations. *Public Relations Review*, 46(2), 101902.

[12] Boisot, M., Canals, A. (2004). Data, Information and knowledge: Have we got it right? *Journal of Evolutionary Economics*, 14, 43–67.

[13] Short, E.J., Todd, S. (2017). What's your data worth? *MIT Sloan Management Review*, 58(3), 17–19.

[14] Evans, P.C., Basole, R.C. (2016). Economic and business dimensions: Revealing the API ecosystem and enterprise strategy via visual analytics. *Communications of the ACM*, 59(2), 26–28.

[15] Tan, W., Fan, Y., Ghoneim, A., Hossain, M.A., Dustdar, S. (2016). From the service-oriented architecture to the web API economy. *IEEE Internet Computing*, 20(4), 64–68.

[16] Zachariadis, M., Ozcan, P. (2017). The API economy and digital transformation in financial services: The case of open banking. *SSRN Electronic Journal. In SWIFT Institute Research Paper Series*.

[17] Davies, T., Perini, F., Alanso, J. (2013). Researching the emerging impacts of open data. World Wide Web Foundation, Washington, DC.

[18] Zeleti, F.A., Ojo, A., Curry, E. (2016). Exploring the economic value of open government data. *Government Information Quarterly*, 33(3), 535–551.

[19] Yoon, H., Zo, H., Ciganek, A.P. (2011). Does XBRL adoption reduce information asymmetry? *Journal of Business Research*, 64(2), 157–163.

[20] Immonen, A., Palviainen, M., Ovaska, E. (2014). Requirements of an open data based business ecosystem. *IEEE Access*, 2, 88–103.

[21] Janssen, M., Zuiderwijk, A. (2014). Infomediary business models for connecting open data providers and

users. *Social Science Computer Review*, 32(5), 694–711.

[22] Zimmermann, H.D., Pucihar, A. (2015). Open innovation, open data and new business models. *SSRN Electronic Journal*, 449–458.

[23] Kitsios, F., Papachristos, N., Kamariotou, M. (2017). Business models for open data ecosystem: Challenges and motivations for entrepreneurship and innovation. IEEE 19th Conference on Business Informatics, 1, 398–407.

[24] Ridder, G., Moffitt, R. (2007). The econometrics of data combination. *Handbook of Econometrics*, 6, 5469–5547.

[25] Hayashi, T., Ohsawa, Y. (2020). Understanding the structural characteristics of data platforms using metadata and a network approach. *IEEE Access*, 8, 35469–35481.

[26] Ohsawa, Y., Kido, H., Hayashi, T., Liu, C. (2013). Data jackets for synthesizing values in the market of data. 17th International Conference in Knowledge Based and Intelligent Information and Engineering Systems, *Procedia Computer Science*, 22, 709–716.

[27] Hayashi, T., Ohsawa, Y. (2018). Retrieval system for data utilization knowledge integrating stakeholders' interests. Beyond Machine Intelligence: Understanding Cognitive Bias and Humanity for Well-Being AI. AAAI Spring symposium.

[28] Ohsawa, Y., Kido, H., Hayashi, T., Liu, C., Komoda, K. (2015). Innovators marketplace on data jackets, for valuating, sharing, and synthesizing data. *Smart Innovation, Systems and Technologies*. Springer, 30, 83–97.

[29] Iwasa, D., Hayashi, T., Ohsawa, Y. (2020). Development and evaluation of a new platform for accelerating cross-domain data exchange and cooperation. *New Generation Computing*, 38, 65–96.

[30] Cao, X., Chen, Y., Liu, K.J.R. (2017). Data trading with multiple owners, collectors, and users: An iterative auction mechanism. *IEEE Transactions on Signal and Information Processing Over Networks*, 3(2), 268–281.

[31] Quix, C., Chakrabarti, A., Kleff, S., Pullmann, J. (2017). Business process modelling for a data exchange platform. *The 29th International Conference on Advanced Information Systems Engineering*, 153–160.

[32] Spiekermann, M. (2019). Data marketplaces: Trends and monetisation of data goods. *Intereconomics*, 54(4), 208–216.